\documentclass[prd,twocolumn,showpacs]{revtex4}

\usepackage{amsmath,amssymb,latexsym}
\usepackage{graphicx}% Include figure files
\usepackage{dcolumn}% Align table columns on decimal point
\usepackage{bm}% bold math

\def\be{\begin{equation}}
\def\ee{\end{equation}}
\newcommand{\bea}{\begin{eqnarray}}
\newcommand{\eea}{\end{eqnarray}}
\newcommand{\nn}{\nonumber}
\def\ie{{\it i.e.}}
\def\eg{{\it e.g.}}
\def\d{{\rm d}}
\def\xL{x_{\rm L}}
\def\rs{r_{\rm S}}
\def\xs{x_{\rm S}}
\def\ys{y_{\rm S}}
\def\ts{t_{\rm S}}
\def\tO{t_{\rm O}}
\def\tL{t_{\rm L}}
\def\dL{d_{\rm L}}
\def\rL{r_{\rm L}}

\begin{document}

\title{A Rigorous Approach to the Gravitational Lensing}
\date{\today}

\author{Minjoon Park}
\affiliation{
Department of Physics, University of California, Davis,
California 95616, USA
}

\begin{abstract}
We revisit a weak gravitational lensing problem 
by constructing a setup which describes the actual system as accurately as possible 
and solving the null geodesic equations. Details are given for the case of a universe
driven only by a cosmological constant, $\Lambda$, which confirm the conventional results: 
The conventional lensing analysis is
correct as it is, without any need for correction of ${\mathcal O}(\Lambda)$.
We also treat the cases of the lensing in generic Friedmann-Robertson-Walker backgrounds.
\end{abstract}


\maketitle

\section{Introduction}
In the conventional analysis on the weak gravitational lensing, the bending of light from a source by 
a massive deflector, or a lens, is calculated in the comoving frame with the Schwarzschild metric, 
whereas the effect of the propagation of light in an expanding universe is taken care of
by using angular diameter distances as the distance measure. The underlying idea for this separative
treatment is that on most of its trajectory a photon does not know the existence of the lens, and 
it feels the gravitational pull only near a small region of the closest approach to the lens where
the spacetime can be locally approximated by Schwarzschild. Although this sounds like a reasonable approximation, 
it is somewhat qualitative, and hard to produce quantitative errors to justify itself. 

Then a better, or at least more precise, description of the situation 
can be given by 1. putting our lens in a Friedmann-Robertson-Walker(FRW) spacetime 
to get the ``Schwarzschild-FRW" spacetime and 
2. solving for the null geodesics connecting a source to an observer in this background.

Step 1 was done about 80 years ago by McVittie\cite{mcvittie}, who got the
exact solutions for the Einstein equation sourced by a localized (spherical) mass and cosmological medium. 
The McVittie solution is given by
\be\label{eqn:mcvittiefull}
\d s^2 = -\Big(\frac{1-\mu}{1+\mu}\Big)^2\d t^2 
+ (1+\mu)^4 a(t)^2 \d \vec X^2 \,,
\ee
where $\vec X$ is the usual comoving coordinate, 
\be
\mu = \frac{m}{4a(t)|\vec X - \vec X_0|} \,, \nn
\ee
with $\vec X_0$ the location of a mass whose Schwarzschild radius is $m$, and the scale factor $a(t)$ 
is a solution of the Friedmann equation without the mass. It can be immediately seen that for a vanishing mass 
(\ref{eqn:mcvittiefull}) is reduced to the FRW metric, 
whereas when there is no cosmological source, \ie, 
$a(t) = 1$, we recover the Schwarzschild metric in isotropic coordinates.
Describing an FRW spacetime with a massive source located 
at fixed comoving coordinates $\vec X_0$, (\ref{eqn:mcvittiefull}) fits well for the analysis of
gravitational lensing where we, an observer at the coordinate origin, watch the deflection of light by
a mass, \ie, a lens, moving away from us according to Hubble's law. 
Then depending on the integrability of the geodesic equations 
we can perform step 2 either analytically or numerically. 

As a clearcut demonstration of our point, we will consider an exactly solvable case of 
a cosmological constant($\Lambda$) driven universe, so that the scale factor is $a(t) = e^{H t}$
with $H = \sqrt{\Lambda/3}$. 
As one can guess naturally, a spacetime with a mass and $\Lambda$ may be able to be described 
by the Schwarzschild-de Sitter(SdS) metric.
Indeed, we can transform (\ref{eqn:mcvittiefull}) into SdS by first transforming 
\be\label{eqn:rtran}
\tilde r = (1+\mu)^2 a(t)\, |\vec X|\,,
\ee
to get
\bea
\d s^2 &=& -\Big(1-\frac{m}{\tilde r} - H^2 \tilde r^2 \Big) \d t^2
+\frac{\d \tilde r^2}{1-m/{\tilde r}} \nn\\
&&- \frac{2H\tilde r}{\sqrt{1-m/{\tilde r}}} \, \d t \d \tilde r
+\tilde r^2 \d \Omega_2\,,
\eea
and then redefining $t$ by $\tilde t = t + f(\tilde r)$ with
\be
\frac{\d f}{\d\tilde r} 
= \frac{H \tilde r}{(1-m/\tilde r-H^2 \tilde r^2)\sqrt{1-m/\tilde r}} \,,
\ee
to reach a familiar form 
\bea\label{eqn:famsds}
\d s^2 &=& -\Big(1-\frac{m}{\tilde r}-H^2 \tilde r^2 \Big) \d \tilde t{}^2 \nn\\
&&+\frac{\d\tilde r^2}{1-\frac{m}{\tilde r}-H^2 \tilde r^2} 
+\tilde r^2\d\Omega_2\,.
\eea

But in the real world, the lens(L), the source(S) and the observer(O) are all
moving according to Hubble's law, and with (\ref{eqn:famsds}) 
it may not be clear how to impose such a requirement, causing interpretational confusion.
Therefore to facilitate the intuitive description of the system
we go back to (\ref{eqn:mcvittiefull}) and define ``physical"
spatial coordinates by
\be\label{eqn:physxdef}
\vec x = e^{H t} \vec X\,,
\ee
because the actual distance is (scale factor)$\times$(comoving distance).
Also, in weak gravitational lensing 
situations, $m$ is much smaller than any other length scale under consideration, 
and we will work only up to ${\cal O}(m)$. Then, the background of our interest is
\begin{widetext}
\be\label{eqn:ourbkg}
\d s^2 = -\Big( 1-\frac{m}{\sqrt{(x+e^{H t}q)^2 + y^2 + z^2}} \Big) \d t^2 
+ \Big( 1+\frac{m}{\sqrt{(x+e^{H t}q)^2 + y^2 + z^2}} \Big) (\d \vec x - H \vec x \,\d t)^2 \,,
\ee
where we align the coordinates to put the lens on the $x$-axis. 
Note that the cosmic expansion is implemented in the coordinates themselves. 
Especially, the Hubble motion of L relative to O can be seen directly from the metric: 
It is moving along $\vec x=(-e^{H t}q,0,0)$, where $q$ is a constant related to the location of the lens 
at a certain moment of time. 

Gravitational lensing in SdS spacetime is not a new subject, and it was shown that $\Lambda$ does not
show up explicitly in observable quantities\cite{islam}: 
For a thin lens, the bending of the photon trajectory occurs for a period much shorter than
$H^{-1}\sim\Lambda^{-1/2}$, so that the spatial expansion due to $\Lambda$ cannot have any effect practically. 
It does affect the propagation of
the photon between S and L, and L and O, but the angular diameter distance properly absorbs it.
But recently \cite{Rindler:2007zz}-\cite{Ishak} took a fresh look at this old subject,
claiming the appearance of ${\mathcal O}(\Lambda)$ correction, and 
there have been pros\cite{Lake:2007dx} and cons\cite{Khriplovich:2008ij} to their claim.
Now we will revisit this problem with the metric of (\ref{eqn:ourbkg}).

\section{Lensing in a cosmological constant driven Universe}
\begin{figure}
  \begin{center}
    \resizebox{12cm}{!}{\includegraphics{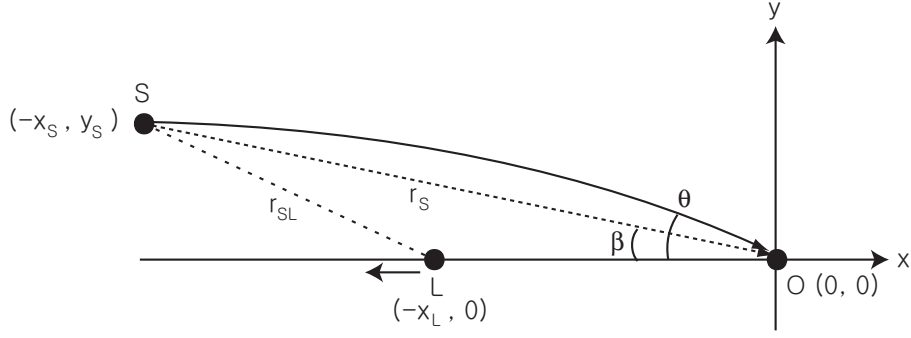}}
  \end{center}
  \caption{Lensing schematics.}
  \label{alg}
\end{figure}
We consider the motion of a photon in the spacetime of (\ref{eqn:ourbkg}) (FIG. \ref{alg}). 
Corresponding Christoffel symbols can be calculated straightforwardly. 
Because of the symmetry of the system, we can confine the photon on the $x$-$y$ plane and set $z=0$. 
Also, the null condition
\be\label{eqn:nullcon}
0 = -\Big( 1 - \frac{m}{R} \Big) + \Big( 1+\frac{m}{R} \Big) \{(x'-H x)^2 + (y'-H y)^2 \} \,,
\ee
with a prime denoting a $t$-derivative and $R=\sqrt{(x + q e^{H t})^2 + y^2}$, can be used to simplify geodesic equations. 
To reduce the number of equations to solve, we parametrize the geodesics by $t$. 
Since $t$ is not an affine parameter, our geodesic equations have the following form:
\be
x^{\mu}{}''+ \Gamma^\mu_{\nu\rho} x^\nu{}' x^\rho{}' = k x^\mu{}'\,.
\ee
With the help of (\ref{eqn:nullcon}), the $t$-geodesic equation determines $k$ to be
\be
k = H + \frac{m}{R^3} \{ -H (x^2 + y^2) + x x' + y y' + e^{H t} q (x'-H x) \} \,,
\ee
and then $x$- and $y$-equations become
\bea\label{eqn:xgeoeq}
0 &=& x'' - H x' + \frac{m}{R^3} \Big\{ H x' (x^2 + y^2) - (x x' + y y') x' + (x y'- x' y) y' 
+ e^{H t} q (H x - x') (H e^{H t} q + 2x') + e^{H t} q \Big\}\,,\nn\\
0 &=& y'' - H y' + \frac{m}{R^3} \Big\{ H y' (x^2 + y^2) - (x x' + y y') y' - (x y'- x' y) x'
+ e^{H t} q (H y - y') (H e^{H t} q + 2x') \Big\}\,.
\eea

Next, we split the photon trajectory into the zeroth order piece(without L) 
and the first order one(with L, of ${\cal O}(m)$):
\be\label{eqn:ansatz}
x = x_0 + \frac{m}{r} x_1\,, \quad y = y_0 + \frac{m}{r} y_1\,,
\ee
where $r$ is the typical length scale of the lens system and $r \gg m$. The zeroth order solution is trivial:
\be\label{eqn:0th}
x_0 = a_1 + a_2 e^{H t} \,, \quad y_0 = b_1 + b_2 e^{H t}\,.
\ee
In the absence of the lens, the boundary conditions are such that at $t = \ts$ 
the photon was at the location of S, 
$(-\xs, \ys)$, and at $t = \tO$ it arrives at O at the origin, $(0,0)$. 
Then, $a_1$, $a_2$, $b_1$, $b_2$ are fixed to be
\bea\label{eqn:abs}
a_1 = -\frac{e^{H \tO}\xs}{e^{H \tO} - e^{H \ts}} \,, \quad
a_2 = \frac{\xs}{e^{H \tO} - e^{H \ts}}\,, \quad
b_1 = \frac{e^{H \tO}\ys}{e^{H \tO} - e^{H \ts}} \,, \quad
b_2 = -\frac{\ys}{e^{H \tO} - e^{H \ts}}\,. 
\eea

Plugging (\ref{eqn:ansatz}) and (\ref{eqn:0th}) into (\ref{eqn:xgeoeq}), 
we get the first order geodesic equations:
\bea\label{eqn:1stxgeoeq}
0 &=& x_1'' - H x_1' + \frac{(a_1^2 + b_1^2)(a_2 + q) + a_1 e^{H t} \{(a_2 + q)^2 + b_2^2\}}
{\{(b_1 + b_2 e^{H t})^2 + (a_1 + e^{H t} (a_2 + q))^2\}^{3/2}} e^{H t} H^2 r \,,\nn\\
0 &=& y_1'' - H y_1' + \frac{(a_1^2 + b_1^2) b_2 + b_1 e^{H t} \{(a_2 + q)^2 + b_2^2 \}}
{\{(b_1 + b_2 e^{H t})^2 + (a_1 + e^{H t} (a_2 + q))^2\}^{3/2}} e^{H t} H^2 r \,,
\eea
whose solutions are
\bea\label{eqn:1st}
x_1 &=& c_2 + c_1 e^{H t} + \frac{r b_2}{q b_1}   
\sqrt{(a_1 + e^{H t}(a_2 + q))^2 + (b_1 + e^{H t} b_2)^2} \nn\\
&&+ \frac{r e^{H t} (a_2 + q)}{\sqrt{a_1^2 + b_1^2}} \log \Big( e^{-H t}H^2
\Big( a_1 (a_1 + e^{H t} (a_2 + q)) + b_1 (b_1 + e^{H t} b_2) \nn\\
&&\qquad\qquad\qquad + \sqrt{a_1^2 + b_1^2} \sqrt{(a_1 + e^{H t} (a_2 + q))^2 + (b_1 + e^{H t} b_2)^2} \;\Big) \Big)\,, \nn\\
y_1 &=& d_2 + d_1 e^{H t} - \frac{r(a_2 + q)}{q b_1} \sqrt{(a_1 + e^{H t} (a_2 + q))^2 + (b_1 +e^{H t} b_2)^2} \\
&& + \frac{r e^{H t} b_2}{\sqrt{a_1^2 +b_1^2}} \log \Big(e^{-H t} H^2 
\Big( a_1 (a_1 + e^{H t}(a_2 + q)) + b_1 (b_1 + e^{H t} b_2) \nn\\
&&\qquad\qquad\qquad + \sqrt{a_1^2 + b_1^2} \sqrt{(a_1 + e^{H t} (a_2 + q))^2 + (b_1 + e^{H t} b_2)^2} \; \Big) \Big) \,. \nn
\eea
When L is present, in order for a photon emitted at the source location to hit O at the origin, 
it should follow a bent trajectory, and the travel time is longer than $\tO - \ts$ 
by an amount of ${\mathcal O}(m)$. Let us say it reaches the origin at $t=\tO+\frac{m}{r}t_1$. 
Then, we need to fix six unknowns, $c_1$, $c_2$, $d_1$, $d_2$, $q$ and $t_1$, to completely determine
the photon trajectory. 

To be a trajectory of a photon (\ref{eqn:ansatz}) must satisfy the null condition: 
Plugging the solutions (\ref{eqn:0th}) and (\ref{eqn:1st}) into (\ref{eqn:nullcon}) gives
\be
0 = H^2 (a_1^2+b_1^2) - 1 + \frac{m}{r}\{A_1(t)(a_1 b_2 - a_2 b_1) 
+ A_2(t)(H^2 (a_1^2+b_1^2) - 1) + A_3(t)(a_1 c_2 + b_1 d_2) \}\,,
\ee
where $A_i$'s are some complicated functions of $t$. Then, at ${\mathcal O}(m^0)$, we get
\be\label{eqn:tr}
H^2 (a_1^2+b_1^2) = 1 \quad \Rightarrow \quad e^{-H(\tO-\ts)} = 1-H \rs\,,
\ee
which relates the source-observer time difference, $\tO - \ts$,  
to the source-observer distance, $\rs= \sqrt{\xs^2 + \ys^2}$, in the absence of the lens.
Next at ${\mathcal O}(m)$ we have 
\be\label{eqn:cdcon}
a_1 c_2 + b_1 d_2 = 0\,,
\ee
and this is our first constraint on the unknowns.
Note that (\ref{eqn:tr}) can be generalized 
by considering (\ref{eqn:nullcon}) at ${\mathcal O}(m^0)$:
\be\label{eqn:trg}
1+Hx = \frac{\d x}{\d t} \quad \Rightarrow \quad 
H(\tO-t_\bullet) = \log(1+Hx_{\rm O}) - \log(1+Hx_\bullet) = - \log(1+Hx_\bullet)\,,
\ee
where $x_\bullet$ is the location of the photon at $t=t_\bullet$. 
The boundary conditions in the presence of the lens give 4 constraints:
\bea\label{eqn:1stxbc}
&x(\ts) = -\xs = -\xs + \frac{m}{r}x_1(\ts)\,, \quad&
x(\tO+\frac{m}{r}t_1) = 0 = x_0'(\tO) \frac{m}{r}t_1 + \frac{m}{r}x_1(\tO)\,,\nn\\
&y(\ts) = \ys = \ys + \frac{m}{r}y_1(\ts)\,, \quad&
y(\tO+\frac{m}{r}t_1) = 0 = y_0'(\tO) \frac{m}{r}t_1 + \frac{m}{r}y_1(\tO)\,.
\eea
The last piece of information necessary for connecting our algebra to the observation is the location of L: 
it was located at $-\rL$ at time $t = \tL$. 
In order for us to detect it at $t=\tO$, $\rL$ and $\tL$
must satisfy (\ref{eqn:trg}), \ie,
\be
e^{-H(\tO-\tL)} = 1-H \rL\,.
\ee 
Then $q$ can be determined by
\be\label{eqn:xL}
-q e^{H\tL} = -\rL \quad \Rightarrow \quad q = e^{-H\tO} \frac{\rL}{1-H\rL} \equiv e^{-H\tO}\xL\,.
\ee
With (\ref{eqn:abs}), (\ref{eqn:tr}) and (\ref{eqn:xL}), solving (\ref{eqn:cdcon}) and (\ref{eqn:1stxbc}) gives
the rest of the unknowns:
\bea\label{eqn:1stcd}
c_1 &=& \frac{r e^{-H \ts}}{H \rs^3 \xL} 
\Big[\, \rs \Big\{-H^2 \rs\xL^2\xs +\Big(H(\rs^2-\xL \xs)-\rs\Big)
\Big(-\xL+\sqrt{(\xs-(1-H \rs) \xL)^2+\ys^2}\;\Big) \Big\} \nn\\
&&\quad-H(1-H\rs)\xL\Big\{\rs^2(H\rs\xL+\xs)\log\frac{e^{-H\ts}H(1-H\rs)(\rs-\xs)\xL}{\rs}\nn\\
&&\qquad + \Big(\rs^2(H\rs\xL+\xs)-\xL\ys^2\Big)
\log \frac{\rs^2-\xL \xs(1-H \rs)+\rs 
\sqrt{(\xs-(1-H \rs)\xL)^2+\ys^2}}{(1-H \rs)(\rs-\xs)\xL}\;\Big\} \,\Big] \,, \nn\\
c_2 &=& \frac{r}{H\rs^3\xL} \Big[\, \rs (\rs+H \xs \xL) 
\Big\{ -\xL(1-H \rs)+\sqrt{(\xs-(1-H \rs)\xL)^2+\ys^2} \;\Big\} \nn\\
&&\quad- H (1-H \rs) \xL^2 \ys^2 
\log \frac{\rs^2-\xL \xs(1-H \rs)+\rs 
\sqrt{(\xs-(1-H \rs)\xL)^2+\ys^2}}{(1-H \rs)(\rs-\xs)\xL} \;\Big] \,, \\
d_1 &=& \frac{r e^{-H\ts}}{H\rs^3 \xL\ys} 
\Big[\, \rs\Big\{H^2\rs\xL^2\ys^2+\Big(H(\rs^2\xs-\xL\xs^2)-\rs\xs+H^2\rs^3\xL\Big)
\Big(-\xL+\sqrt{(\xs-(1-H \rs)\xL)^2+\ys^2}\;\Big)\Big\} \nn\\
&&\quad+ H(1-H\rs)\xL\ys^2\Big\{\rs^2\log\frac{e^{-H\ts}H(1-H\rs)(\rs-\xs)\xL}{\rs} \nn\\
&&\qquad + (\rs^2+\xL\xs)\log \frac{\rs^2-\xL \xs(1-H \rs)+\rs 
\sqrt{(\xs-(1-H \rs)\xL)^2+\ys^2}}{(1-H \rs)(\rs-\xs)\xL} \;\Big\}\Big] \,, \nn\\
d_2 &=& - \frac{a_1}{b_1} c_2\,, \nn
\eea
and 
\bea\label{eqn:1stt}
t_1 &=& \frac{r}{1-H \rs} \Big\{ H \Big(-(1-H \rs) \xL 
+ \sqrt{(\xs-(1-H \rs)\xL)^2+\ys^2} \;\Big)  \nn\\
&&+ (1-H \rs) \Big(1+H \xL \frac{\xs}{\rs}\Big)
\log \frac{\rs^2-\xL \xs(1-H \rs)+\rs 
\sqrt{(\xs-(1-H \rs)\xL)^2+\ys^2}}{(1-H \rs)(\rs-\xs)\xL} \;\Big\} \,.
\eea
With (\ref{eqn:1stcd}) and (\ref{eqn:1stt}), we can see that $x_1$, $y_1$ and $t_1$ 
have an overall factor of $r$, so that the arbitrary parameter $r$ does not appear in the full solution.
$t_1$ has a special implication: $\frac{m}{r}t_1$ is related to the {\it time delay} due to the lens. 

Now that the photon trajectory is determined,
we can calculate the angular location of the image of S observed by us:
\bea\label{eqn:theta}
\theta &=& -\tan^{-1}\frac{y'}{x'}\Big|_{\,\tO+\frac{m}{r}t_1}\nn\\
&=& \beta + \frac{m}{\xL \ys}
\Big\{ \xs + H \rs \xL + \Big(1+H \xL\frac{\xs}{\rs} \Big)
\Big(-(1-H \rs)\xL + \sqrt{(\xs-(1-H \rs)\xL)^2+\ys^2} \;\Big) \nn\\
&&\qquad\qquad - \frac{\ys^2}{\rs^2} H \xL^2 (1-H \rs)
\log \frac{\rs^2-\xL \xs(1-H \rs)+\rs 
\sqrt{(\xs-(1-H \rs)\xL)^2+\ys^2}}{(1-H \rs)(\rs-\xs)\xL} \,\Big\} 
+ {\mathcal O}(m^2) \,,
\eea
where $\beta = \tan^{-1} \frac{\ys}{\xs}$ is the undeflected image location, and $\xL=\rL/(1-H\rL)$.
This is the \emph{exact} result up to ${\mathcal O}(m)$. 

\vspace{10pt}
\begin{figure}
  \begin{center}
    \resizebox{10cm}{!}{\includegraphics{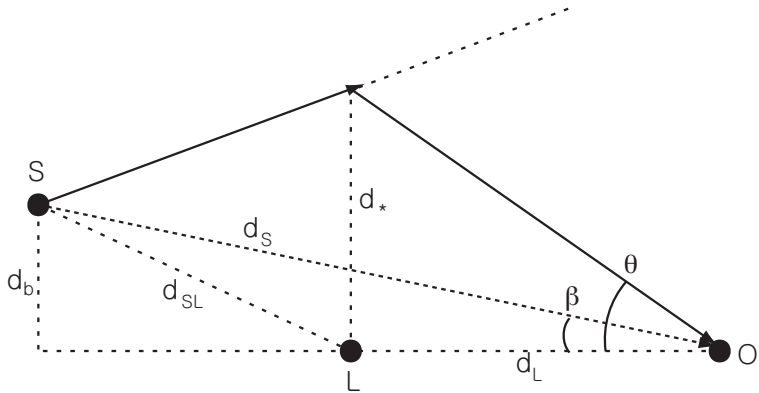}}
  \end{center}
  \caption{Lensing schematics in conventional analysis}
  \label{phys}
\end{figure}

On the other hand the conventional lensing analysis(FIG. \ref{phys}), 
\eg, \S~IV of \cite{Keeton:2005jd}, gives
\be\label{eqn:contheta}
\theta_c \approx \beta_c + \frac{2m d_{\rm SL}}{\beta_c \dL d_{\rm S}} + {\cal O}(m^2) \,,
\ee
where $d_\bullet$'s are angular diameter distances defined by
$d_\bullet = a(t_\bullet) \rho_\bullet$
with $\rho_\bullet$ being the corresponding comoving distance. Obviously
\be
\beta_c \approx \frac{d_b}{d_{\rm S}}\,,\quad\theta_c \approx \frac{d_*}{\dL} \,.
\ee
To compare (\ref{eqn:contheta}) to our result (\ref{eqn:theta}), 
we first note that our $\xs$, $\ys$ and $\rs$ are
distances at $t=\ts$ and $\rL$ at $t=\tL$. Then from (\ref{eqn:physxdef}),
we immediately see that these are the same as the angular diameter distances, \ie,
$\rs=d_{\rm S}$, $\rL=\dL$, 
and therefore $\beta_c = \beta$, $\theta_c = \theta$.
Now for small $H$, (\ref{eqn:theta}) becomes
\be\label{eqn:apptheta}
\theta = \beta + \frac{2m}{\beta d_{\rm S}\dL} 
\Big\{ \xs-\dL+H \dL(\xs-\dL) + H^2 \dL^2(\xs-\dL) 
+ {\mathcal O}(H^3) + {\mathcal O}(\beta^2) \Big\} + {\mathcal O}(m^2) \,.
\ee
Using
\bea\label{eqn:dsl}
d_{\rm SL} &=& a(\ts)\sqrt{\Big(\frac{\xs}{a(\ts)}-\frac{\dL}{a(\tL)}\Big)^2+\Big(\frac{\ys}{a(\ts)}\Big)^2} 
= \sqrt{\Big(\xs - \frac{1-H d_{\rm s}}{1-H\dL}\dL  \Big)^2 + \ys^2}\nn\\
&=& \xs-\dL+H \dL(\xs-\dL) + H^2 \dL^2 (\xs-\dL) 
+ {\mathcal O}(H^3) + {\mathcal O}(\beta^2)\,,
\eea
we can rewrite (\ref{eqn:apptheta}) as
\be\label{eqn:conthetaconv}
\theta = \beta + \frac{2m d_{\rm SL}}{\beta d_{\rm S} \dL}
\Big( 1 + {\cal O}(H^3) + {\cal O}(\beta^2) \Big) + {\cal O}(m^2) \,,
\ee
where 
\be
{\cal O}(\beta^2) = -\beta^2 \frac{\xs^2-4\xs\dL+2\dL^2}{4(\xs-\dL)} + \cdots\,.
\ee
Therefore, our result is in contradiction to the recent claims by \cite{Rindler:2007zz}
which assert that there should be a ${\mathcal O}(\Lambda) \sim {\mathcal O}(H^2)$ correction 
to the conventional lensing analysis. 
\end{widetext}

\section{Discussion}
Through rigorous, close-to-the-reality but laborious derivations, we confirm that
the conventional gravitational lensing analysis, although it may look loosely constructed, is actually
accurate, at least for thin, weak lenses in a $\Lambda$ driven universe. 
\cite{Khriplovich:2008ij}, which incorporated the ``reality" in their analysis by employing 
FRW coordinates, has the same conclusion as ours.

The differences between our work and \cite{Rindler:2007zz}-\cite{Ishak}(and \cite{Lake:2007dx})
can originate from the following: In \cite{Rindler:2007zz},
\begin{enumerate}
\item
the setup is not describing the actual observation: 
In the SdS universe, L and O must be moving relative to each other due to Hubble expansion, 
whereas they got an angle measured by an observer with fixed coordinates,
\item
their result is not written in terms of angular diameter distances, 
which are necessary to compare with the conventional results.
\end{enumerate}
In their follow-up paper \cite{Ishak}, some effort was made to take these into consideration. But
\begin{itemize}
\item
their method to resolve problem 2, \ie, employing the Einstein-Strauss scheme or the lens equation, 
does not seem to be used properly: They are necessary in the Schwarzschild lensing 
because the Hubble expansion, which is not considered in the calculation of the bending angle, 
has to be implemented. On the other hand, the SdS spacetime is a complete playground 
for gravitational lensing because the metric is equivalent to the McVittie solution 
which already knows about both the lens and FRW. 
Then appropriate specifications of O, L and S should be enough for the full description of phenomena. 
Obviously, using Einstein-Strauss scheme or the lens equation together with SdS background is redundant.
\item
the problem 1 was not addressed: Their angle $\psi$(in the first paper) 
or $\alpha$(in the second paper) is measured by a static observer(O). 
The instantaneous speed $v$ of a geodesically moving observer(O$\prime$) can be calculated to be
\be
v = \sqrt{\frac{m}{r_L} + H^2 r_L^2}\,,
\ee
where $r_L$ is the radial coordinate of O(O$\prime$) at the moment of the observation. 
Then, by relativistic aberration, the angle observed by O$\prime$ is
\be\label{eqn:psi'}
\psi' = \sqrt{\frac{1+v}{1-v}} \psi\,,
\ee
assuming $\psi(\psi')$ is small. To get the final result, we should convert 
quantities appearing in (\ref{eqn:psi'}) into angular diameter distances. This last step could be
tricky, which makes the intuitive setup presented in the previous section favorable. 
\end{itemize}

The same rigorous procedure can be applied to generic FRW systems. For example, in the case of matter domination
with $a(t) \sim t^{2/3}$, once finishing tremendously more complicated intermediate steps, 
we can write the final result as
\begin{widetext}
\be\label{eqn:appthetainmdu}
\theta = \beta + \frac{2m}{\beta d_{\rm S}\dL} 
\Big\{ \xs-\dL+\frac{2}{3\tO} \dL(\xs-\dL) 
+ \Big(\frac{2}{3\tO}\Big)^2 \dL(\xs-\dL) \frac{3\xs+7\dL}{4}
+ {\mathcal O}\Big(\frac{2}{3\tO}\Big)^3 + {\mathcal O}(\beta^2) \Big\} 
+ {\mathcal O}(m^2) \,.
\ee
\end{widetext}
After relating $d_{\rm SL}$ to other variables in a similar way as (\ref{eqn:dsl}), we get
\be
\theta = \beta + \frac{2md_{\rm SL}}{\beta d_{\rm S}\dL} 
\Big( 1 + {\mathcal O}(H_{\rm O}^3) + {\mathcal O}(\beta^2) \Big) + {\mathcal O}(m^2) \,,
\ee
with $H_{\rm O} = 2/(3\tO)$, which still ascertains the accuracy of the conventional analysis.

In addition to confirming the conventional results, our work can provide the exact formulae
for errors coming from dropping higher order terms of ${\mathcal O}(H_{\rm O}^3)$ and ${\mathcal O}(\beta^2)$,
which might have importance in testing theories of long-range modification of gravity: 
Since our results are obtained under the assumption that the general relativity(GR) is valid up to any scale, 
if precise measurements of high redshift lensing systems differ from our prediction, 
it may indicate that GR needs some modifications at large distances.
At present, these theoretical errors are well buried beneath the measurement ones, 
but as our observational tools develop they may become significant in the future.

\section*{Acknowledgement}
We thank C. Fassnacht, A. Iglesias, C. Keeton and especially N. Kaloper for discussions.

\end{document}